\documentstyle[mprocl,psfig]{article}

\def\Journal#1#2#3#4{{#1} {\bf #2}, #3 (#4)}

\def\PRD{{\em Phys. Rev.} D}

\def\be{\begin{equation}}
\def\ee{\end{equation}}
\def\bea{\begin{eqnarray}}
\def\eea{\end{eqnarray}}

\begin{document}

\title{Three Dimensional Distorted Black Holes: Initial Data and Evolution}

\author{ S. R. Brandt, K. Camarda, E. Seidel }
\address{
Max-Planck-Institut f\"ur Gravitationsphysik\\
Schlaatzweg 1, 14473 Potsdam, Germany
}
\maketitle\abstracts{
We present a new class of 3D black hole initial data sets for numerical
relativity.  These data sets go beyond the axisymmetric, ``gravity wave plus
rotating black hole'' single black hole data sets by creating
a dynamic, distorted hole with adjustable distortion parameters in 3D.
These data sets extend our existing test beds for 3D numerical relativity,
representing the late stages of binary black hole collisions resulting from
on-axis collision or 3D spiralling coalescence, and should provide insight
into the physics of such systems.  We describe the construction of these
sets, the properties for a number of example cases, and report on progress
evolving them.
}
\section{Introduction}
In all that follows, the standard ADM 3+1 decomposition is used\cite{ADM} along with the
standard York decomposition of the initial value problem\cite{YorkIVP}.  This work
extends previous studies of axisymmetric black holes\cite{BernsteinIVP,brandt} to
allow for a dependence on the azimuthal coordinate.  The basic idea is to combine
the pure gravitational wave construction due to Brill and combine it with a black
hole to form adjustable, highly distorted black hole initial data sets.

\section{Initial Value Problem I:  Non-Axisymmetric Distorted Black Holes}
The first type of solution we consider for a non-axisymmetric gravity wave is the
Brill wave form superimposed on an Einstein-Rosen bridge,
used by 2D simulations at NCSA\cite{BernsteinIVP}.  In these
time-symmetric spacetimes, the three metric is given by
\begin{equation}
ds^2 = \psi^4 \left(e^{2 q}\left[d\eta^2+d\theta^2\right]+\sin^2\theta\,d\phi^2\right).
\end{equation}
In this equation, $q$ may be chosen freely (so long as the appropriate boundary conditions are
respected), but the form we use here is similar to that used by previous NCSA work\cite{BernsteinIVP}
\begin{eqnarray}
q &=& Q_0 \sin^n\theta \left(1+c \cos^2\phi\right)\tilde{g}\\
\tilde{g} &=&
e^{-\left(\eta-\eta_0\right)^2/\sigma^2}+
e^{-\left(\eta+\eta_0\right)^2/\sigma^2}.
\end{eqnarray}
In the above $n$ is an even integer, constrained such that $n \ge 2$, which specifies the angular
dependence of the Brill wave.  The parameter $\sigma$ specifies its width, and $\eta_0$
its radial position.
The parameter $c$ specifies the non-axisymmetric aspect of the wave, and 
when $c=0$ we recover the axisymmetric case\cite{BernsteinIVP}.
Once these parameters are supplied, we numerically solve the Hamiltonian constraint equation
for $\psi$, the conformal factor.
The initial value code was tested against the results of the 2D code for accuracy, and
convergence studies show that it converges to second order.

\section{Initial Value Problem II:  Non-Axisymmetric Distorted Black Holes with Angular Momentum}
Another form of distorted black hole initial data,
one which allows us to have both a non-axisymmetric gravity wave and
angular momentum, is based on a conformally flat space ($q=0$).
Here, the gravity wave
is constructed by specifying the conformal extrinsic curvature.
Its parameters are
an angular momentum $J$, a $\phi$-dependence given by $m$ (normally
we use either $m=4$ so that the IVP is isometric under
the operation $\phi \rightarrow \phi \pm \pi/2$,
or we use $m=0$ to facilitate comparison with 2D codes),
and a function of $\eta$ that is typically chosen to be the same as $\tilde{g}$ above, and
is arbitrary so long as it vanishes as $\eta\rightarrow\infty$ and is symmetric at $\eta=0$.
Here we sketch the procedure of constructing the solution to the constraints:
writing $K_{ij}=\psi^{-2} \hat{K}_{ij}$
we take
\begin{eqnarray}
\hat{K}_{ij} &=& m \sin(m\phi) \left[\begin{array}{ccc}
h_0 & h_1 & 0 \\
h_1 & h_x-h_0 & 0 \\
0 & 0 & -h_x \sin^2\theta \end{array}\right]
+ \left[\begin{array}{ccc}
0 & 0 & \hat{K}_{\eta\phi} \\
0 & 0 & \hat{K}_{\theta\phi} \\
\hat{K}_{\eta\phi} & \hat{K}_{\theta\phi} & 0
\end{array}\right] \\
\hat{K}_{\eta\phi} &=& 
\sin^2\theta \left(3 J+\cos\left(m\phi\right)
\frac{1}{\sin^3\theta}\partial_\theta \left(\sin^4\theta g\right)\right)\\
\hat{K}_{\theta\phi} &=& \cos\left(m\phi\right)
\sin\theta \left(m^2 v+\sin^2\theta \partial_\eta g\right).
\end{eqnarray}
If we now make the substitutions:
\begin{eqnarray}
h_0 &=& \partial_\eta \Omega-\frac{1}{\sin\theta} \partial_\theta \left( \sin\theta \Lambda\right)\\
h_1 &=& \partial_\eta \Lambda+\frac{1}{\sin\theta} \partial_\theta \left( \sin\theta \Omega\right)
\end{eqnarray}
then the $\eta$- and $\theta$-momentum constraints become:
\begin{eqnarray}
\left(\partial_\eta^2-1+\partial_\theta^2+2\cot\theta\partial_\theta\right)\Omega &=&
-\left(4 \cos\theta + \sin\theta \partial_\theta \right) \partial_\phi g\\
\left(\partial_\eta^2-1+\partial_\theta^2+2\cot\theta\partial_\theta\right)\Lambda &=&
-\csc\theta \partial_\phi v-\partial_\theta h_x+\sin\theta \partial_{\eta}\partial_\phi g
\end{eqnarray}
Carefully choosing the forms and boundary
values of the functions on the right hand side will then allow us to solve the constraints
analytically in terms of the function $\tilde{g}$ and its derivatives.
Details will be published in a future paper.
\section{Evolutions of type I initial data}
To evolve this data, we first interpolate it from the spheroidal $(\eta,\theta,\phi)$ coordinate
system to a Cartesian coordinate system used by our evolution codes.
For the present we limit our evolutions to a grid with an
outer boundary placed at $18 M$, so the results we obtain should not be taken as more than
a feasibility study.  Higher resolution runs,
on much larger grids, are in progress and will be published\cite{CamardaPhD,EdAndK} elsewhere.
Nevertheless, we are able to evolve both axisymmetric data ($c=0$),
and non-axisymmetric data ($c \neq 0$) and obtain qualitatively similar results.  The metric elements
evolve in the same fashion, with the same qualitative features we are familiar with from 2D codes.
Below we show two figures.  In Figure I we see the growth of the metric function for a run parameterized
by $(Q_0 = -0.1, \eta_0 = 0.0, \sigma = 1.0, n=4, c=0.5)$. 
The growth of the metric function here is familiar from all
1D and 2D studies performed by the group.  In Figure II we see the beginnings of a waveform
extracted at several radii and can watch its progress as it travels outward.  The wavelength
matches well with the value predicted from linear theory, despite the nearness of the outer
boundary.

\begin{tabular}{cc}
\psfig{figure=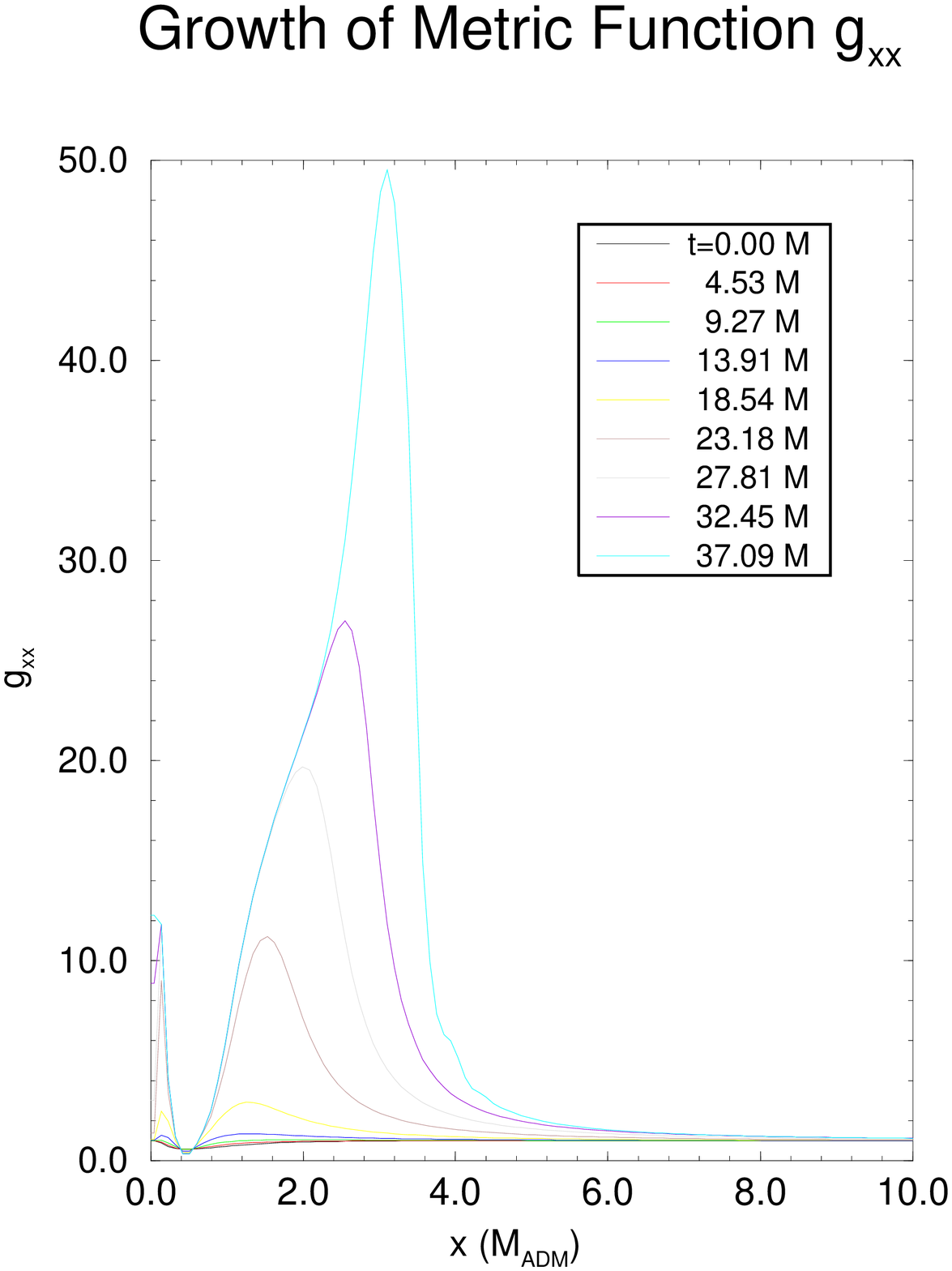,height=6cm,width=5.0cm} &
\psfig{file=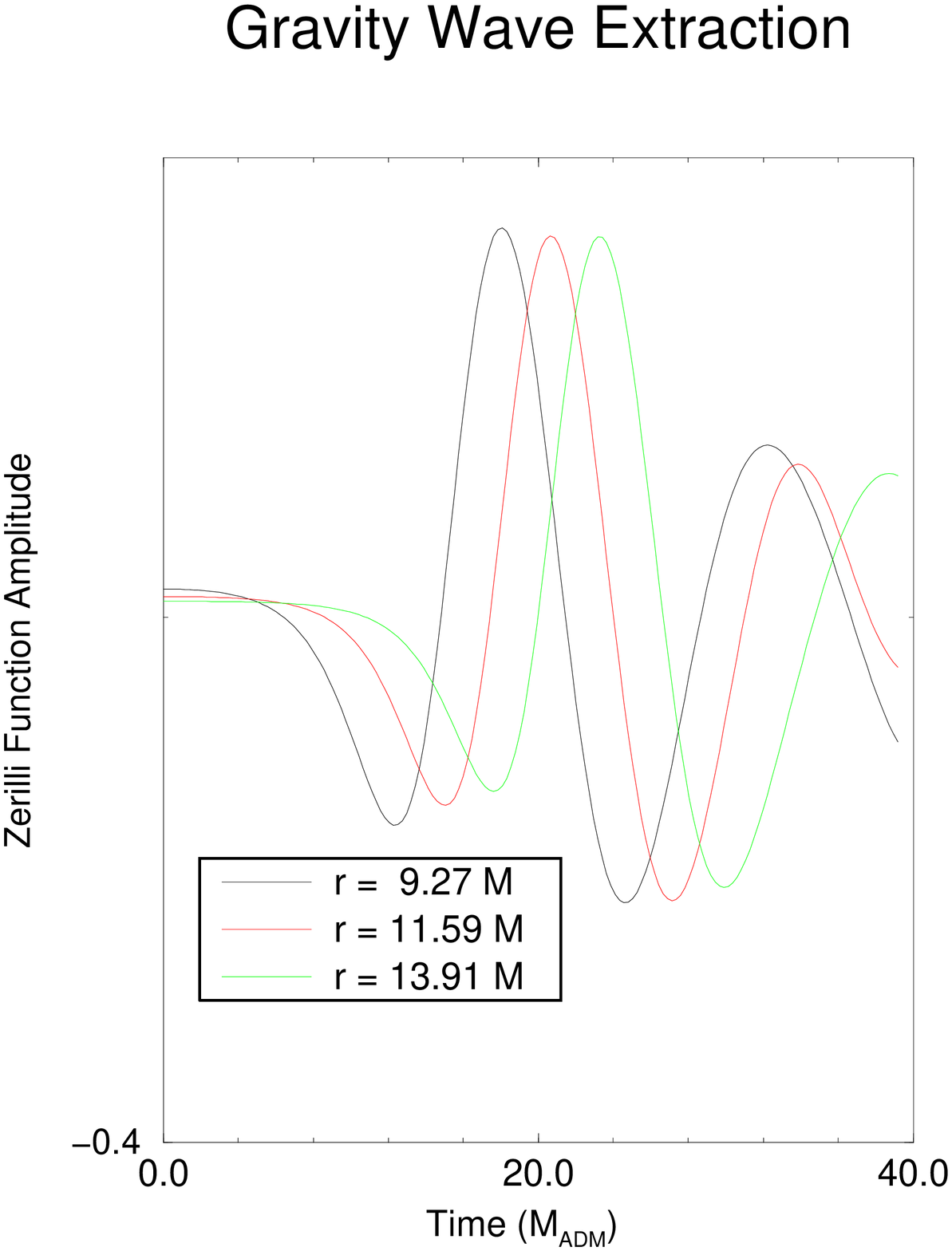,height=6cm,width=5.0cm} \\
\begin{minipage}[t]{6 cm}
Figure 1: Metric function $g_{xx}$ evaluated along the x-axis and plotted at
several times.
\end{minipage}&
\begin{minipage}[t]{6cm}
Figure 2: The $\ell=2$ Zerilli function extracted at three different radii and
plotted as a function of time.  The wave function is clearly propagating outward.
\end{minipage}
\end{tabular}

\section{Future Work}

As of this writing, much of this work has already been extended\cite{CamardaPhD}.
Furthermore, in the near future we plan to publish full descriptions of these data sets, their
implementations, and to describe in detail comparisons with 2D code results (where
the outer boundary may be placed arbitrarily far away), and convergence tests.

\end{document}